\documentclass[11pt]{amsart}
\usepackage{amsmath}
\usepackage{amsfonts}
\usepackage{amssymb}
\usepackage{amsthm}

\newcommand{\chan}{\mathcal E}

\newcommand{\dual}[1]{#1^*}
\newcommand{\chanh}{\dual\chan}

\newcommand{\bra}[1]{\langle #1 |}
\newcommand{\ket}[1]{| #1 \rangle}

\newcommand{\proj}[1]{\ket{#1}\bra{#1}}
\newcommand{\tr}{\text{Tr}}
\newcommand{\one}{{\bf 1}}
\newcommand{\Hil}{\mathcal H}

\newcommand{\be}{\begin{equation}}
\newcommand{\ee}{\end{equation}}

\newcommand{\finops}{\mathcal M}

\newcommand{\id}{{\mathrm{id}}}

\theoremstyle{plain}
\newtheorem{prop}{Proposition}[section]
\newtheorem{thm}[prop]{Theorem}

\newtheorem{cor}[prop]{Corollary}

\theoremstyle{definition}
\newtheorem{dfn}{Definition}[section]

\newcommand{\ind}[1]{#1}

\begin{document}

\title[Quantum error correction]{Quantum error correction on infinite-dimensional Hilbert spaces}
\author[C. B\'eny]{C\'edric B\'eny$^{1,2}$}
\address{$^1$Centre for Quantum Technologies, National University of Singapore, 3 Science Drive 2, Singapore 117543}
\address{$^2$Departments of Applied Mathematics and Physics, University of Waterloo, ON, Canada, N2L 3G1}
\author[A. Kempf]{Achim Kempf$^{2,4}$}
\author[D.W. Kribs]{David~W.~Kribs$^{3,4}$}
\address{$^3$Department of Mathematics and Statistics, University of
Guelph, Guelph, ON, Canada, N1G 2W1} \address{$^4$Institute for
Quantum Computing, University of Waterloo, ON Canada, N2L 3G1}
\date{\today}

\begin{abstract}
We present a generalization of quantum error correction to infinite-dimensional Hilbert spaces. The generalization yields new classes of quantum error correcting codes that have no finite-dimensional counterparts. The error correction theory we develop begins with a shift of focus from states to algebras of observables. Standard subspace codes and subsystem codes are seen as the special case of algebras of observables given by finite-dimensional von Neumann factors of type I. Our generalization allows for the correction of codes characterized by any von Neumann algebra and we give examples, in particular, of codes defined by infinite-dimensional algebras.
\end{abstract}

\maketitle


\section{Introduction}

A common challenge in the numerous fields of quantum information
science is to devise techniques that protect the evolution of
quantum systems from external perturbations. Since interactions
with the environment are generally unavoidable, a variety of
so-called quantum error correction procedures have been developed
in order to correct the effect of environmental noise on quantum
systems. The basic idea underlying the correction procedures is to
exploit any knowledge that one may possess about the nature of the
noise. To this end, the quantum information is usually encoded
into a larger system in such a way that interactions with the
environment affect essentially only the auxiliary degrees of
freedom. In this way, the original qubits remain retrievable;
i.e., the errors are correctable. So far, research in quantum
error correction has been concerned primarily with the special
case of a finite number of qubits which are embedded in a
finite-dimensional Hilbert space. Real systems are, of course,
always ultimately described in an infinite-dimensional Hilbert
space. Algorithms for the cases of a finite or infinite number of
qubits encoded in an infinite-dimensional Hilbert space have been
proposed, in particular see~\cite{braunstein98x1, lloyd98,
gottesman01, braunstein05} respectively. However, no general
theory (for instance, in the spirit of ~\cite{knill97}) exists.

Here, we lay the foundations for quantum error correction in
infinite-dimensional Hilbert spaces in full generality. We find
that many of the basic results for quantum error correction extend
to the infinite-dimensional setting, and that there are new
phenomena which appear only in the infinite-dimensional setting.
The new phenomena extend beyond obvious features such as the
possibility of an infinite number of individual errors. In
particular, we uncover types of infinite-dimensional codes that
have no finite-dimensional counterparts or approximations.

Historically, following the realization that quantum error
correction is possible and the discovery of seminal
examples~\cite{bennett96, shor95, steane96, gottesman96}, Knill
and Laflamme found a general mathematical condition characterizing
the codes that are correctable for a given arbitrary noise
model~\cite{knill97}. In this framework, the state of the
information to be corrected is encoded in a subspace of the
(finite-dimensional) physical Hilbert space. It was later
realized~\cite{knill00, kribs05, kribs06} that there is a more
subtle way of encoding the relevant information, namely in a {\em
subsystem}. This amounts to assuming that we ignore the effect of
the noise on the complementary subsystem. In finite-dimensions,
this does not lead to larger quantum codes, but to more efficient
correction procedures~\cite{poulin05x1, bacon06, aliferis07,
aly07}. It was shown in~\cite{beny07x1, beny07x4} that this idea
can be further generalized if, instead of a subsystem, one focuses
on a restricted set of observables, which, for the purpose of
quantum error correction, can always be assumed to span a
finite-dimensional algebra. We will show that this idea naturally
generalizes to infinite-dimensional Hilbert spaces, where the
finite-dimensional algebras are replaced by arbitrary von Neumann
algebras. In this context, a subsystem, as defined by a
tensor-product structure on the underlying Hilbert space,
corresponds to a von Neumann factor of type I. Other types of
factors, however, also correspond to full-fledged logical quantum
systems.

This paper is organized as follows. In the next section we give
relevant background. We then begin our investigation, which
includes an analysis of the sharp fixed and correctable
observables; a derivation of the simultaneously correctable
observables and operator systems; a brief review of von Neumann
algebras and discussion of the special cases captured by the
theory; presentation of explicit type I and type II examples; and
finally an outlook section.

\section{Background} \label{section:qec}

\subsection{Quantum error correction} \label{subsection:qec}

We begin with a general review of quantum error correction
motivated by the presentation of \cite{knill00}, which focused on
the finite-dimensional case but equally well applies to the case
of bounded interaction operators (as described below). Full
details on our formalism in the infinite-dimensional setting will
be provided below. Let us suppose that some information is sent
through a quantum channel $\chan$. The aim of quantum error
correction is to find certain degrees of freedom, the {\em error
correcting code}\index{error correcting code} (whose exact nature
we deliberately keep imprecise for the moment), on which the
effect of the channel can be inverted. Since the inversion must be
implemented physically, it must be a valid physical
transformation; i.e., a channel. The inverse channel $\mathcal R$
is called the {\em \ind{correction channel}}. Under reasonable
assumptions (see Section \ref{subsection:formalism}) a channel
$\chan$ can always be written as
\[
\chan(\rho) = \sum_i E_i \rho E_i^\dagger,
\]
which means that we can assume the noise is given by a discrete family of individual error operators $E_i$ on Hilbert space.

In fact we will see that if $\mathcal R$ corrects this channel on
some code, then it will correct also any channel whose elements span
the same operator space as the elements of $\chan$. This  is
important because often one does not know the precise channel
elements $E_i$. Indeed, suppose that the system interacts
continuously with its environment via a general Hamiltonian $H =
H^\dagger$ of the form
\begin{equation}
H = \sum_i J_i \otimes K_i
\end{equation}
where the {\em interaction operators} $J_i$ act on the system
Hilbert space, and the operators $K_i$ act on the environment.

In order for the state of the system at a given time to depend
unambiguously on its initial state, we must assume that the initial
state of the environment is uncorrelated with that of the system.
For simplicity, we will further assume that it is a pure state
$\ket{\psi}$. It is unlikely that we know much about $K_i$ or
$\ket{\psi}$ given that the environment may be very large and
complex. However, it is generally conceivable that we have good
knowledge of what the operators $J_i$ can be. If $\chan_t$
is the channel describing the evolution of the system alone up to
time $t$, and $\mathcal{E}_t^*$ is the corresponding dual map, then
we have
\begin{equation}
\begin{split}
\chanh_t(A) &= (\one \otimes \bra{\psi}) e^{it H} (A \otimes \one)e^{-it H}(\one \otimes \ket{\psi})\\
 &= \sum_k (\one \otimes \bra{\psi}) e^{it H}(\one \otimes \ket{k}) A (\one \otimes \bra{k})e^{-it H}(\one \otimes \ket{\psi})\\
&= \sum_k E_k^\dagger(t) A E_k(t)
\end{split}
\end{equation}
where the channel elements are:
\begin{equation}
\label{equ:EJ}
\begin{split}
E_k(t) &= (\one \otimes \bra{k})e^{-it H}(\one \otimes \ket{\psi})\\
&= \sum_n \frac{(-it)^n}{n!} (\one \otimes \bra{k}) H^n (\one \otimes \ket{\psi})\\
&= \sum_n \sum_{j_1 \dots j_n} \frac{(-it)^n}{n!} \bra{k} K_{j_1} \cdots K_{j_n}\ket{\psi} J_{j_1}\cdots J_{j_n}.  \\
\end{split}
\end{equation}
Hence, we know that no matter what the environment operators and initial state are, the span of the channel elements $E_k(t)$ belongs to the algebra generated by the interaction operators $J_i$. One can look for correctable codes for all channels with this property.
Such codes would be a form of
infinite-distance code (in fact an example of noiseless subsystems
\cite{knill00}). While there are important instances in which
these codes do exist, they form a rather restrictive class of
quantum error correcting code.

On the other hand, if the time $t$ at which we aim to perform the
correction is small enough, we can do better. Indeed, suppose that
$\lambda$ is some interaction parameter with unit of energy, then
the above series is expressed in powers of $t\lambda$. We see that
to the $n$th order in $t \lambda$, the elements of the channel are
in the span of the $n$th order products $J_{j_1}\cdots J_{j_n}$.
Hence, if we correct often enough (in order to limit the value of
$t \lambda$), then we only need to find a correction channel and a
code for channel elements in the span of the operators
$J_{j_1}\cdots J_{j_n}$, $n < N$, for a fixed $N$. In this
context, the operators $J_i$ are seen as representing individual
errors, and our code corrects up to $N$ independent errors
\cite{knill00}.

For clarity of the presentation we will stick to the simple
picture where the channel $\chan$ is given. However we will keep
the more realistic situation in mind and check that the correction
procedure we devised works not just for the given channel, but
also for any channel whose elements span the same space.

\subsection{Stochastic Heisenberg picture}


Traditionally, one attempts to correct {\em states}, namely to
simultaneously find a state $\rho$ and a channel $\mathcal R$ such
that $\mathcal R(\chan(\rho)) = \rho$. In \cite{beny07x1,
beny07x4}, it was shown that it is convenient to consider instead
the correction of {\em observables}. Most generally, an observable
is specified by a positive operator valued measure (POVM). For
instance, if the measure is discrete, the POVM $X$ is specified by
a family of positive operators $X_k$ called {\em effects}. In
general an effect can be any positive operator smaller than the
identity: $0 \le X_k \le \one$. In order to form a discrete POVM,
these effects must sum to the identity: $\sum_k X_k = \one$. More
generally, a POVM $X$ sends measurable subsets of a measure space
$\Omega$ to positive operators. It is defined by the effects
$X(\omega)$, $\omega \subseteq \Omega$. For a discrete POVM,
$X(\omega) = \sum_{k \in \omega} X_k$.

What matters physically are the expectation values of the form
\[
\tr(\rho A)
\]
where $A$ is an effect. Indeed, for a POVM $X$, $\tr(\rho
X(\omega))$ is the probability that the outcome of a measurement
of $X$ falls inside $\omega \subseteq \Omega$.

If the channel $\chan$ describes the evolution of our system, a
measurement of an observable $X$ after the evolution will yield
probabilities of the form $\tr(\chan(\rho) X(\omega))$.
Alternatively, one can define a dual map $\chanh$ which satisfies
\[
\tr(\chan(\rho) X(\omega)) = \tr(\rho \,\chanh(X(\omega))
\]
for all $\omega \subseteq \Omega$. This dual channel is the stochastic
form of the usual Heisenberg picture of quantum mechanics. It specifies
the evolution of observables instead of states.
In this picture, an observable $X$ evolves to the observable $Y$
defined by $Y(\omega) = \chanh(X(\omega))$. An important
conceptual point to note here, which is not apparent in the usual
case of unitary evolutions, is that this evolution goes
``backward'' in time. Indeed, if the channel $\chan_2$ follows the
channel $\chan_1$ in time, the overall transformation in the
Schr\"odinger picture is given by the channel $\chan_2 \circ
\chan_1$. But the dual channels compose in the reverse order; we
have $(\chan_2 \circ \chan_1)^* = \chan_1^* \circ \chan_2^*$. This
means that the latest transformation must be applied first.

Instead of correcting states, we may thus attempt to correct
observables; i.e., to pair a channel $\mathcal R$ with an
observable $X$ such that
\[
(\mathcal R \circ \chan)^*(X(\omega)) = X(\omega)
\]
for all $\omega \subseteq \Omega$. This expression means that
measuring $X$ before or after the action of the map $\mathcal R \circ \chan$
would yield the same outcomes with the same probabilities no
matter what the initial state was. Hence we can say that the observable
$X$ is corrected by $\mathcal R$ for the noise defined by $\chan$.
In general, if the channel $\mathcal R$ exists such that this
equation is satisfied, we say that the observable $X$ is {\em
correctable}. The sets of simultaneously correctable {\em sharp}
observables, that is correctable by a single common channel
$\mathcal R$, were characterized in \cite{beny07x1} and shown to
generalize all previous known types of quantum error correcting
codes. We will show here that this approach leads naturally to an
infinite-dimensional generalization.


\subsection{States and channels in infinite dimensions} \label{subsection:formalism}

We consider a quantum system characterized by a Hilbert space $\Hil$
that may be infinite-di\-mensional.
For effects, it is natural to consider the {\em bounded} linear
operators on $\Hil$. Indeed, the condition $0 \le A \le \one$
guarantees that $A$ is a bounded operator on $\Hil$. We denote the
set of bounded operators on $\Hil$ by $\mathcal B(\Hil)$, and the
set of effects (or positive contractions on $\Hil$) by
$\mathbb{E}(\Hil)$. The set $\mathcal B(\Hil)$ is naturally a von
Neumann algebra. In general we could assume that states are
positive linear functionals on $\mathcal B(\Hil)$. In the finite
dimensional case, this guarantees that they are of the form $A
\mapsto \tr(\rho A)$ for some operator $\rho$ and all effects $A$.
However, this does not carry through to the infinite-dimensional
case. Instead, we postulate that a state is represented by a
positive operator $\rho$ such that $\tr(\rho A)$ is well defined
for any effect $A$, and also such that $\tr(\rho) = 1$. This means
that states are {\em trace-class} operators. Formally, the
trace-class operators $B \in \mathcal B(\Hil)$ are those for which
the expression
\[
\sum_i \bra{i} \sqrt{B^\dagger B} \ket{i}
\]
converges for (one and hence) any basis $\ket{i}$. We will let
$\mathcal{B}_t(\Hil)$ denote the set of trace-class operators on
$\Hil$. For self-adjoint elements $\rho \in \mathcal B_t(\Hil)$,
we have a {\em \ind{trace}} defined by
\[
\tr(\rho) := \sum_i \bra{i} \rho \ket{i}.
\]
The product of an element of $\rho \in \mathcal B_t(\Hil)$ with
any operator $A \in \mathcal B(\Hil)$ is also trace-class, which
implies that we can define $\tr(\rho A)$. The set $\mathcal
B_t(\Hil)$ itself is a Banach algebra. It is the pre-dual of
$\mathcal B(\Hil)$, which means that $\mathcal B(\Hil)$ is the set
of linear functionals on $B_t(\Hil)$. This means that effectively
we have defined the effects as linear functionals on states rather
than the converse. The existence of a pre-dual is a fundamental
property of von Neumann algebras.

A general von Neumann algebra is equipped with a \ind{weak-$*$
topology}, which amounts to defining the convergence of a sequence
$A_n \in \mathcal B(\Hil)$ in terms of expectation values. In the
case of $\mathcal B(\Hil)$, this means that the sequence converges
to $A$ if and only if the numbers $\tr(\rho A_n)$ converge to
$\tr(\rho A)$ as $n \rightarrow \infty$ for all states $\rho \in
\mathcal B_t(\Hil)$.

This implies that the states represented by elements of $\mathcal
B_t(\Hil)$, seen as linear functionals of effects, are continuous
with respect to the weak-$*$ topology. Indeed, if the sequence
$\{A_n\}_{n=1}^\infty$ converges to $A$ in this topology, then, by
definition $\tr(\rho A_n) \rightarrow \tr(\rho A)$. Hence the map
$A \mapsto \tr(\rho A)$ is continuous. Conversely, those are {\em
all} the weak-$*$ continuous positive linear functionals.
Therefore, our choice of states corresponds to restricting the
natural set of all linear functionals on effects to only those
which are weak-$*$ continuous, or {\em normal}\index{normal
states} for short.

A channel from a system represented by the Hilbert space $\Hil_1$
to a system represented by $\Hil_2$ can be defined by a
trace-preserving completely positive linear map
\[
\chan: \mathcal B_t(\Hil_1) \rightarrow \mathcal B_t(\Hil_2)
\]
on states. It has a dual
\[
\chanh: \mathcal B(\Hil_2) \rightarrow \mathcal B(\Hil_1)
\]
which describes the evolution of effects, and hence observables,
in the Heisenberg picture. A channel can be represented as
\[
\chanh(A) = \sum_{k=1}^\infty E_k^\dagger A E_k
\]
or
\[
\chan(\rho) = \sum_{k=1}^\infty E_k \rho E_k^\dagger
\]
where the sum can now be infinite \cite{kraus83}. We will call
this the
operator-sum form of the channel $\chan$. The elements $E_k$ are
bounded linear operators from $\Hil_1$ to $\Hil_2$. This can be
understood starting from the \ind{Stinespring dilation} theorem
for completely positive maps between C$^*$-algebras, which states
that there is a representation $\pi$ of $\mathcal B(\Hil_2)$ on
some Hilbert space $\mathcal{K}$, and an isometry $V: \Hil_1
\rightarrow \mathcal{K}$, such that
\[
\chanh(A) = V^\dagger \pi(A) V
\]
for all $A \in \mathcal B(\Hil_2)$. In the case that we are
considering (von Neumann factor of type I, and normal map), the
representation $\pi$ on $\Hil$ is of the form $\pi(A) = A \otimes
\one$. Also, if $\Hil_2$ is separable, then so is $\mathcal{K}$
\cite{paulsen02}. Therefore the subsystem on which $\pi(A)$ acts
trivially is also separable, and possesses a discrete basis
$\ket{i}$. This implies that
\[
\begin{split}
\chanh(A) &= V^\dagger (A \otimes \one) V = \sum_i V^\dagger (A \otimes \proj{i}) V  \\
&= \sum_i (\, V^\dagger \ket{i} \, )A  (\, \bra{i}  V \, ),
\end{split}
\]
where the operators $\bra{i} V : \Hil_1 \rightarrow \Hil_2$ are
defined by the induced tensor structure on the dilation space
$\mathcal{K}$. Thus the elements of the channel $\chan$ can be
chosen to be $E_i = \bra{i}\, V$. Note from this observation it is
clear there is a large ambiguity in the choice of the elements
$E_i$. Indeed, any orthonormal basis $\ket{i}$ would potentially
yield a different set of elements.

\section{Sharp Correctable Observables}

A POVM $X$ is correctable if the effect of the channel can be
inverted on all the effects $X(\omega)$ in the following sense.
\begin{dfn}\label{correctdfn}
We say that an effect $0 \le A \le \one$ is {\em correctable} for
the channel $\chan$ if there exists a channel $\mathcal R$ such
that
\[
A = \chanh(\mathcal R^*(A))
\]
\end{dfn}
In this section we will characterize the correctability of certain
types of effects, namely projectors. This is important because the
effects of a {\em sharp observable}; i.e., a traditional
observable represented by a self-adjoint operator, are always
projectors (they are the spectral projectors of the corresponding
self-adjoint operator). The results obtained in this section will
form the basis of our understanding of correctable observables. We
begin with a ``warm up'', the case of passive error correction in
this setting.

\subsection{Sharp fixed observables}

We consider the problem of characterizing the sharp observables
that are unaffected by the action of the channel; that is, which
are correctable in the above sense but with the trivial correction
channel $\mathcal R(\rho) = \rho$. This requires that we use the
same source and destination Hilbert spaces, namely
\[
\Hil := \Hil_1 = \Hil_2.
\]
\begin{dfn}
We say that an observable $X$ is {\em fixed} by the channel
$\chan$ if
\[
X(\omega) = \chanh(X(\omega)) \quad \text{for all $\omega$}.
\]
\end{dfn}
A slightly more general form of this problem (see below) was addressed for channels defined on
finite-dimensional Hilbert spaces in~\cite{beny07x1} and shown to
yield all noiseless subsystems~\cite{knill00, holbrook03, choi05}.
These results may be readily generalized to the
infinite-dimensional setting, provided that we model the proof on
the approach of~\cite{beny07x4} which does not refer to the
structure theory of finite-dimensional algebras.

Since we focus on sharp observables, the effects $X(\omega)$ are
all projectors. Let us therefore first characterize the fixed
projectors $P$, which satisfy $P = \chanh(P)$. By multiplying on
both sides by the orthogonal projector $P^\perp = \one - P$, we
obtain $0 = P^\perp \chanh(P) P^\perp = \sum_{k} P^\perp
E_k^\dagger P E_k P^\perp$. The right hand side is a sum of
positive operators, which must therefore all equal zero: $(P E_k
P^\perp)^\dagger P E_k P^\perp = 0$ for all $k$, which in turns
implies
\begin{equation}
\label{equ:ns1} P E_k P^\perp = 0
\end{equation}
for all $k$. Similarly, because the dual channel is always unital,
we have $P^\perp = \chanh(P^\perp)$, from which we also deduce
\begin{equation}
\label{equ:ns2} P^\perp E_k P = 0.
\end{equation}
Combining Eq. (\ref{equ:ns1}) and Eq. (\ref{equ:ns2}), we obtain
$P E_k = E_k P$ for all $k$. Hence the fixed projectors must
commute with all the channel elements. This condition is in fact
sufficient since it implies
\[
\chanh(P) = \chanh(\one) P = P.
\]
Hence we have proved the following.
\begin{prop}
\label{lemma:ns} A projector $P$ satisfies $\chanh(P) = P$ if and
only if it commutes with every channel element for $\chanh$;
\begin{equation}
\label{equ:ns} [P, E_k] = 0 \quad \text{ for all $k$.}
\end{equation}
\end{prop}

We see that the commutant algebra of the channel elements, given
by
\[
\mathcal A^{N} := \{ A \in \mathcal B(\Hil) : [A, E_k] =
[A^\dagger, E_k] = 0 \; \forall k\},
\]
which is a von Neumann algebra, plays a central role here. Indeed,
since a von Neumann algebra is spanned by its projectors,
$\mathcal A^{N}$ is spanned by all the fixed projectors. In
addition, the channel clearly fixes all the elements of $\mathcal
A^{N}$. We will call $\mathcal A^{N}$ the {\em noiseless algebra}
for $\chan$.

Since a sharp observable $X$ is fixed by definition when all of
its elements $X(\omega)$ are fixed, we have from
Proposition~\ref{lemma:ns} that it is fixed if and only if its
spectral projectors $X(\omega)$ belong to $\mathcal A^{N}$, which
is equivalent to simply asking that the corresponding self-adjoint
operator $\widehat X$ belongs to $\mathcal A^{N}$.
\begin{prop}
\label{prop:ns}
A sharp observable $X$ represented by the self-adjoint operator
$\widehat X$ is fixed by $\chan$ if and only if $\widehat X \in
\mathcal A^{N}$; i.e.,
\[
[\widehat X, E_k] = 0 \quad \text{for all $k$}.
\]
\end{prop}

We will show in Section \ref{section:oqec}, in the more general
setting were the correction procedure may be non-trivial, that
when the algebra $\mathcal A^{N}$ is a factor of type I, it
corresponds to a noiseless subsystem. All types of factors can
emerge in this way. Type II and III factors define new types of
noiseless subsystems which have no finite-dimensional counterpart.
An example of a noiseless factor of type II will be given in
Section \ref{section:extypetwo}.

Let us remark that we can obtain more noiseless algebras by considering the sharp
observables which are fixed only provided a certain restriction on
states characterized by a subspace $\Hil_0 \subseteq \Hil$.
Let us introduce the isometry
\[
V: \Hil_0 \rightarrow \Hil
\]
which embeds $\Hil_0$ into $\Hil$, that is, $V^\dagger V = \one_{\Hil_0}$ and $V V^\dagger$ is the projector of $\Hil$ onto $\Hil_0$. We write $\chan_0$ for the channel $\chan$ restricted to the states in the subspace $\Hil_0$,
\[
\chan_0(\rho) := \chan(V \rho V^\dagger).
\]
This channel has elements $E_k V$. We cannot directly apply Proposition \ref{prop:ns} because this channel does not have the same source and destination spaces. In order to define what it means for an observable of $\Hil_0$ to be fixed by $\chan_0$ we have to specify how we will map back the output of $\chan_0$ from $\Hil$ to $\Hil_0$. The most natural way to do this is to apply the dual of the isometry. Other possibilities would correspond to different type of ``corrections'', and would enter the more general framework presented in the next section.

Hence we must characterize the projectors $P$ satisfying
\[
\chanh_0(V P V^\dagger) = P.
\]
Note that the map $X \mapsto \chanh_0(V^\dagger X V)$ is not unital so that we cannot apply Proposition \ref{prop:ns} directly to it either. However, a small variation of the steps followed in the proof of Proposition \ref{prop:ns} yields the necessary and sufficient condition
\[
(V P V^\dagger) E_k V = E_k V P.
\]
We leave the details to the interested reader.
The noiseless algebra here is
\[
\begin{split}
\mathcal A^{N}_0 = \{ A \in \mathcal B(\Hil_0) : \;& V A V^\dagger E_k V = E_k V A, \\
&V A^\dagger V^\dagger E_k V = E_k V A^\dagger  \; \forall k \}.
\end{split}
\]
Thus each subspace of the Hilbert space is associated with a
different noiseless algebra.

If $\Hil$ is finite-dimensional, the noiseless algebras that are
factors correspond precisely to the noiseless
subsystems~\cite{beny07x1}. The problem in the characterization of
noiseless subsystems for finite-dimensional Hilbert spaces lies
with the identification of the subspaces which support a
nontrivial noiseless algebra.



\subsection{Sharp correctable observables}

We move now to the case of arbitrary correction operations. If $A$
is an effect such that Definition~\ref{correctdfn} holds, then
there exists an effect $B$ such that $A = \chanh(B)$. Let us
consider the correctable sharp effects. A sharp effect $P \in
\mathcal B(\Hil_1)$ is a an effect which is also a projection; $P^2 =
P = P^\dagger$. If it is correctable, then there is an effect $B
\in \mathcal B(\Hil_2)$ such that
\begin{equation}
\label{equ:preseff} P = \chanh(B).
\end{equation}
In order to characterize the correctable sharp effects, we want to
obtain an equivalent condition which involves only the channel
elements $E_k$. To do this,
we multiply Eq.~(\ref{equ:preseff}) from both sides by $P^\perp =
\one - P$ to obtain
\[
P^\perp \chanh(B) P^\perp = P^\perp P P^\perp = 0.
\]
This implies $B E_k (\one - P) = 0$ for all $k$; i.e.,
\begin{equation}
\label{equ:preseff1} B E_k = B E_k P.
\end{equation}
There is another similar equation that we can use. Indeed,
$\chanh(\one - B) = \one - \chanh(B) = \one - P$, and $\one - B$
is a valid effect. Using the same trick as above, we obtain $(\one
- B) E_k P = 0$; i.e.,
\begin{equation}
\label{equ:preseff2} E_k P = B E_k P.
\end{equation}
Combining Eq.~(\ref{equ:preseff1}) and Eq.~(\ref{equ:preseff2}),
we get
\begin{equation}
\label{equ:preseff3} B E_k = E_k P.
\end{equation}
This means than the existence of an effect $B$ which satisfies $B
E_k = E_k P$ for all channel elements $E_k$ is a necessary
condition for $P$ to be correctable. Note that this condition also
implies $\chanh(B) = P$, since
\[
\chanh(B) = \sum_k E_k^\dagger B E_k = \sum_k E_k^\dagger E_k P =
\chanh(\one) P = P.
\]

Thus we have obtained a condition that depends on the channel
elements $E_k$ independently rather than on the whole channel.
However, a much more useful characterization of the correctable
effects requires the elimination of any reference to the unknown
operator $B$. First note that by taking the adjoint of
Eq.~(\ref{equ:preseff3}) we obtain $E_k^\dagger B = P
E_k^\dagger$. Together with Eq.~(\ref{equ:preseff3}), this implies
\[
E_k^\dagger E_j P = E_k^\dagger B E_j = P E_k^\dagger E_j.
\]
This is a necessary condition for $P$ to be correctable which
does not involve the unknown effect $B$. Algebraically, this
condition is precisely the one found in \cite{beny07x1} for the
finite-dimensional case. In the following result we show this
condition is also sufficient in the infinite-dimensional case,
which requires a more delicate analysis.

\begin{thm}
\label{thm:preseff} A sharp effect $P$ is correctable for the
channel $\chan(\rho) = \sum_k E_k \rho E_k^\dagger$ if and only if
\begin{equation}
\label{equ:preseff4} [P, E_i^\dagger E_j] = 0 \text{ for all
$i$,$j$.}
\end{equation}
\end{thm}
\proof We have already proven the necessity. In order to prove the
sufficiency, we will use this condition to build the effect $B$ of
which $P$ is the image. We will need the completely positive map
$\chan_\lambda$ defined by
\begin{equation}
\label{equ:chanlambda} \chan_\lambda(A) := \sum_{i=0}^\infty
\lambda_i E_i A E_i^\dagger
\end{equation}
where $\lambda_i := 2^{-i}$. This choice of $\lambda$ guarantees
that the sum converges in norm for any effect $A$ since $\|E_i A
E_i^\dagger\| \le 1$. In fact, any choice of components $\lambda_i
> 0$ which makes this sum weak-$*$ convergent for any effect $A$
would be sufficient for our purpose. Note that if $\lambda_i = 1$
for all $i$, then $\chan_\lambda = \chan$, which is defined only
on trace-class operators.

We will try the ansatz
\begin{equation}
\label{equ:presansatz} B := (\chan_\lambda(\one))^{-1}
\chan_\lambda(P).
\end{equation}
First, we have to show that $\chan_\lambda(\one)$ can indeed be
inverted. Note that $\chan_\lambda(\one) \ket{\psi} = 0$ if and
only if $\bra{\psi} \chan_\lambda(\one) \ket{\psi} = 0$ since it
is positive. If we write the channel in terms of the elements
$E_k$, we obtain a sum of positive terms which must all be equal
to zero: $\bra{\psi} E_i E_i^\dagger \ket{\psi} = 0$. Hence
$E_i^\dagger \ket{\psi} = 0$ for all $i$. This means that if
$\ket{\psi}$ is in the kernel of $\chan_\lambda(\one)$, it must be
in the kernel of each $E_i^\dagger$, and therefore be orthogonal
to the range of each $E_i$. This shows that we can invert
$\chan_\lambda(\one)$ on the range of any of the operators $E_i$.
In the following we will always be able to assume that
$(\chan_\lambda(\one))^{-1}$ operates on the span of the ranges of
the operators $E_i$. In particular, Eq.~(\ref{equ:presansatz}) is
well-defined.

Now that we have defined $B$, let us check that $\chanh(B) = P$.
We have
\[
\begin{split}
B E_i &= (\chan_\lambda(\one))^{-1} \chan_\lambda(P) E_i \\
&= (\chan_\lambda(\one))^{-1} \sum_k \lambda_k E_k P E_k^\dagger E_i \\
&= (\chan_\lambda(\one))^{-1} \sum_k \lambda_k E_k E_k^\dagger E_i P  \\
&= (\chan_\lambda(\one))^{-1} \chan_\lambda(\one) E_i P  = E_i P,  \\
\end{split}
\]
which is Eq.~(\ref{equ:preseff3}) and proves that $\chanh(B) = P$.
However we have to check that $B$ is an effect (i.e. a positive
contraction). Note that $B E_i =
E_i P$ implies $\sum_i \lambda_i B E_iE_i^\dagger = \sum_i
\lambda_i E_i P E_i^\dagger$. In other words  $B
\chan_\lambda(\one) = \chan_\lambda(P)$, which in turn yields $B =
\chan_\lambda(P) (\chan_\lambda(\one))^{-1}$.
 From the definition of $B$, we also have $B = (\chan_\lambda(\one))^{-1} \chan_\lambda(P)$. Hence we have $[(\chan_\lambda(\one))^{-1}, \chan_\lambda(P)] = 0$, which implies that
\begin{equation}
\label{equ:almostR} B = \chan_\lambda(\one)^{-\frac{1}{2}}
\chan_\lambda(P) \chan_\lambda(\one)^{-\frac{1}{2}} \ge 0
\end{equation}
In addition, $B \le \chan_\lambda(\one)^{-\frac{1}{2}} \one
\chan_\lambda(\one)^{-\frac{1}{2}} = \one$, which shows that $B$
is an effect. Moreover, $B$ is obtained from $P$ by the map
$\mathcal R^*$ defined by
\[
\mathcal R^*(A) = \chan_\lambda(\one)^{-\frac{1}{2}}
\chan_\lambda(A) \chan_\lambda(\one)^{-\frac{1}{2}}.
\]
This map is positive by construction, and we have already shown
that it is unital. It is the dual of the quantum channel
\begin{equation}
\label{equ:R} \mathcal R(\rho) =
\chanh_\lambda(\chan_\lambda(\one)^{-\frac{1}{2}} \, \rho \,
\chan_\lambda(\one)^{-\frac{1}{2}}),
\end{equation}
which therefore is the correction channel for all correctable
sharp observables. \qed

If $\Hil_1$ is finite-dimensional, one can use $\lambda_i = 1$. In
this case, Eq.~(\ref{equ:R}) reduces to the explicit way of
writing the correction channel used in \cite{beny07x1}. This
explicit form has also appeared in \cite{blume-kohout08}.

Theorem~\ref{thm:preseff} also yields some unsharp correctable
effects. Indeed, given the linearity of $\chanh$, if two effects
are corrected by the same channel $\mathcal R$, then so is any of
their convex combinations (which are also effects). Therefore, the
convex hull of the correctable projectors is entirely correctable.
In fact, the continuity of this channel implies that the weak-$*$
closure of this convex hull is also correctable.

Consider the commutant $\mathcal A$ of the operators $E_i^\dagger
E_j$; i.e., the set of operators that commute with each of them,
\[
\mathcal A := \{ A : [A, E_i^\dagger E_j] = 0 \text{ $\forall$
$i$, $j$} \}.
\]
This set is clearly an algebra. In fact, it is a von Neumann
algebra, which always has the property that the set of effects it
contains is the closed convex hull of its projectors
\cite{davidson96}. Since all the projectors in $\mathcal A$ are
corrected by $\mathcal R$, then so are all the effects it
contains. This proves the following.
\begin{cor}
\label{cor:coralg} The set of effects spanning the von Neumann
algebra
\begin{equation}
\label{equ:corralg} \mathcal A = \{A \in \mathcal B(\Hil_1) \; :
\; [A,E_i^\dagger E_j] = 0 \text{ for all $i$,$j$}\}
\end{equation}
are all corrected by the channel defined in Eq.~(\ref{equ:R}). In
addition, this algebra contains all the correctable sharp effects.
\end{cor}

We now formalize the notion of a correctable positive
operator-valued measure.
\begin{dfn}
We say that a POVM $X$ is {\em correctable}\index{correctable
observable} for $\chan$ if there is a channel $\mathcal R$, called
the {\em \ind{correction channel}}, which is such that
\[
X(\omega) = \chanh(\mathcal R^*(X(\omega)))
\]
for all $\omega \subseteq \Omega$. We then say that $X$ is {\em
fixed}\index{fixed observable} by the channel $\mathcal R \circ
\chan$.
\end{dfn}
The operational meaning of this definition is clear. If $X$ is
fixed by $\mathcal R \circ \chan$, then a measurement of $X$ after
the application of the channel $\chan$ followed by $\mathcal R$
will yield the same outcome as if nothing had happened to the
system.

Since the effects associated with a sharp observable are all
projectors, Corollary \ref{cor:coralg} tells us that the
correctable sharp observables are precisely those which are
represented by a self-adjoint operator in the algebra $\mathcal
A$.
\begin{thm}
\label{thm:corrcond} A sharp observable $X$, represented by the
self-adjoint operator $\widehat X$, is correctable for $\chan$ if
and only if \( \widehat X \in \mathcal A. \)
\end{thm}

\section{Simultaneously Correctable Observables and Operator Systems}
\label{section:simultcorr}

We know that, in fact, sharp correctable observables can all be
corrected with the same correction channel.
Let us explicitly define this concept.
\begin{dfn}
A set of POVMs is said to be {\em simultaneously correctable} for
$\chan$ if every POVM in the set is correctable by the same
correction channel $\mathcal R$, which means that they are all
fixed by $\mathcal R \circ \chan$.
\end{dfn}
Such a set is what constitutes our most general notion of a {\em
quantum error correcting code.} These simultaneously correctable
observables represent a consistent set of properties of the system
which can be saved from the noise. However, instead of working
with a set of POVMs, it is more convenient to work with their
effects.

Clearly, if a set of effects is simultaneously correctable, then
so are all linear combinations which are also effects. Since, in
addition, effects are self-adjoint, this means that a set of
simultaneously correctable effects will always be characterized by
an {\em \ind{operator system}} \ind{$\mathcal S$}; i.e., a linear
subspace $\mathcal S \subseteq \mathcal B(\Hil_1)$ which contains
the adjoint of all its elements~\cite{paulsen02}. Hence, if we
say that an operator system $\mathcal S$ is
correctable\index{correctable operator system}, we mean that all
the effects in it are correctable. In fact, given that we are
considering normal channels, it follows that any weak-$*$ limit of
such linear combination will be correctable by the same
correction channel. Hence we shall consider weak-$*$ closed
operator systems.

Therefore our most general notion of a quantum code is a weak-$*$
closed operator system generated by a set of simultaneously
correctable effects $\mathcal S$. Clearly, $\mathcal S$ represents
a set of simultaneously correctable observables, namely all the
observables $X$ with effects $X(\omega) \in \mathcal S$ for all
$\omega$.

Our task is to characterize the maximal correctable operator
systems for a channel $\mathcal{E}$. From the previous section we
already know one: the algebra $\mathcal A$ defined in
Eq.~(\ref{equ:corralg}).
\begin{prop}
\label{prop:corralg} The POVMs with effects in $\mathcal A$ are
all simultaneously correctable by the channel defined in
Eq.~(\ref{equ:R}).
\end{prop}

We will call $\mathcal A$ {\em the \ind{correctable algebra}} for
$\chan$. This is justified by the fact that any other algebra
spanned by correctable effects is inside $\mathcal A$. Indeed,
such an algebra would be spanned by its projectors, but $\mathcal
A$ already contains all the correctable projectors.

We will now show that many other correctable operator systems can
be obtained via Proposition \ref{prop:corralg}. Consider any
subspace $\Hil_0 \subseteq \Hil_1$, and let
\[
V : \Hil_0 \rightarrow \Hil_1
\]
be the isometry which embeds $\Hil_0$ into $\Hil_1$. This means
that the operator $P_0 := V V^\dagger$, defined on $\Hil_1$, is
the projector on the subspace $\Hil_0$, whereas $V^\dagger V =
\one_0$ is the identity inside $\Hil_0$. We can define a new
channel $\chan_0$ by restricting the channel $\chan$ to the
subspace $\Hil_0$, which, physically, amounts to making sure that
the initial state is prepared inside $\Hil_0$. Hence we define
\[
\chan_0(\rho) := \chan( V\rho V^\dagger)
\]
whose dual is
\[
\chanh_0(A) =  V^\dagger \chanh( A ) V
\]
This channel has its own correctable algebra:
\[
\mathcal A_0 = \{ A \in \mathcal B(\Hil_0) \; : \; [A, V^\dagger
E_k^\dagger E_l V] = 0 \}.
\]
This algebra can be naturally embedded in $\mathcal B(\Hil_1)$ via
the isometry $V$; the map $A \mapsto V A V^\dagger$ from $\mathcal
B(\Hil_0)$ to $\mathcal B(\Hil_1)$ is a normal $*$-homomorphism.
Note however that the identity on $\mathcal B(\Hil_0)$ is sent to
the projector $VV^\dagger = P$.

The algebra $V \mathcal A_0 V^\dagger$ may or may not be a
subalgebra of the correctable algebra $\mathcal A$. If it is not,
then its effects are not correctable for the channel $\chan$
itself. However, we will see that it is one-to-one with a family
of simultaneously correctable effects which do not form an
algebra. Indeed, let $\mathcal R_0$ be the correction channel for
$\chan_0$, which is a channel from $\mathcal B_t(\Hil_2)$ to
$\mathcal B_t(\Hil_0)$, and define the set
\begin{equation}
\label{equ:simcorrsys} \mathcal S_0 := \chanh(\mathcal
R_0^*(\mathcal A_0)).
\end{equation}
Note that $\mathcal S_0$ not equal to $\mathcal A_0$ because we
have used $\chanh$ instead of $\chanh_0$. In fact, we have
\begin{equation}
\mathcal A_0 = V^\dagger \mathcal S_0 V.
\end{equation}
This subset $\mathcal S_0 \subset \mathcal B(\Hil_1)$ is in
general not an algebra, but it is always an operator system. We
claim that all the effects which are in $\mathcal S_0$ are
simultaneously correctable. Indeed, for any effect $A =
\chanh(\mathcal R_0^*(B)) \in \mathcal S_0$, where $B \in \mathcal
A_0$, we have
\[
\begin{split}
\chanh(\mathcal R_0^*(V^\dagger A V)) &= \chanh(\mathcal R_0^*(V^\dagger \chanh(\mathcal R_0^*(B)) V))\\
 &= \chanh(\mathcal R_0^*(\chanh_0(\mathcal R_0^*(B))))\\
 &= \chanh(\mathcal R_0^*(B)) = A.
\end{split}
\]
Hence, the correction map is
\[
\rho \mapsto V \mathcal R_0 (\rho) V^\dagger ,
\]
which is a valid channel from $\mathcal B_t(\Hil_2)$ to $\mathcal
B_t(\Hil_1)$.

This shows that, for any subspace $\Hil_0 \subseteq \Hil_1$ we can construct
an operator system $S_0$ whose effects are all simultaneously
correctable. In addition, it is clear that all the observables
which are formed from effects in $S_0$ are simultaneously
correctable. We have thus established the following result.
\begin{thm}
\label{thm:corrsys} For every subspace $\Hil_0 \subseteq \Hil_1$,
the operator system $\mathcal S_0$ defined in
Eq.~(\ref{equ:simcorrsys}) is such that all the POVMs $X$ with
$X(\omega) \in \mathcal S_0$ for all $\omega$ are simultaneously
correctable.
\end{thm}
Let us summarize how to obtain $\mathcal S_0$. We restricted the
channel to the subspace $\Hil_0$, computed the correctable algebra
$\mathcal A_0$ and the correction channel $\mathcal R_0$ for the
restricted channel, and finally set $\mathcal S_0 =
\chanh(\mathcal R_0^*(\mathcal A_0))$. In fact we can be entirely
explicit. Letting $V$ be the isometry embedding $\Hil_0$ into
$\Hil_1$, $P := VV^\dagger$ the projector on $\Hil_0$,
$\chan_\lambda$ the regularized channel defined on the whole of
$\mathcal B(\Hil_1)$, and
\[
K := (\chan_\lambda(P))^{-\frac{1}{2}}
\]
we have that the operator system
\[
\begin{split}
\mathcal S_0 = \{ \chanh(K \chan_\lambda(VAV^\dagger) K)\; : \; & A \in \mathcal B(\Hil_0), \\
&[A, V^\dagger E_i^\dagger E_j V] = 0 \text{ for all $i$, $j$}\}
\end{split}
\]
is corrected on $\chan$ by the channel
\[
\mathcal R(\rho) = P \chan_\lambda^*(K \rho K) P.
\]
For an explicit example in finite-dimension, see \cite{beny07x4}.

Do these structures exhaust all the correctable observables for
$\chan$? Theorem 3 of \cite{blume-kohout08} states that, when
$\Hil_1$ is finite-dimensional, the fixed point set of the dual of
any channel must be made of elements of the form $A + \mathcal
F^*(A)$ where $A$ belongs to a $*$-algebra $\mathcal A_0$ inside
$\finops(\Hil_0)$, $\Hil_0$ a subspace of $\Hil_1$, and $\mathcal
F^*$ is a fixed channel which is such that $P \mathcal F^*(A) P =
0$, where $P$ projects onto $\Hil_0$.

If an operator system $\mathcal S$ is correctable for $\chan$,
then it is fixed by $\chanh \circ \mathcal R^*$ for some channel
$\mathcal R$, and, according to \cite{blume-kohout08}, made of
elements of the form mentioned above: $A + \mathcal F^*(A)$. This
means that if $V$ is the isometry embedding $\Hil_0$ into
$\Hil_1$, we have $V^\dagger \chanh(\mathcal R^*(A+\mathcal
F^*(A))) V = V^\dagger (A+\mathcal F^*(A))V = A$, which implies
that the algebra $\mathcal A_0$ is correctable for $\chan$
restricted to $\Hil_0$, and therefore belongs to the correctable
algebra for this restricted channel. This shows that $\mathcal S$
is of the form covered by Theorem \ref{thm:corrsys}, which
therefore exhausts all correctable observables on a
finite dimensional system.

\subsection{Simultaneously correctable channels}
\label{section:varchan}

In Subsection \ref{subsection:qec} we motivated a situation where, of
the channel elements $E_i$, only their span is known. It is
already clear that the correctable algebra
\[
\mathcal A = \{ A \; : \; [A,E_i^\dagger E_j] = 0 \text{ for all
$i$, $j$} \}
\]
only depends on the span of the elements $E_i$. Indeed, consider a
channel $\chan$ with elements $E_i$ and a channel $\mathcal E'$
with elements $F_i = \sum_j \gamma_{ij} E_j$ where $\gamma_{ij}$
are arbitrary, provided that $\sum_i F_i^\dagger F_i = \one$. Then
it is clear that an operator which commutes with the products
$E_i^\dagger E_j$ for all $i$ and $j$ will also commute with
operators $F_i^\dagger F_j$ since they are just linear
combinations of the former. This fact tells us that $\mathcal A$
is the correctable algebra for all channels whose elements are
chosen in the span of the operators $E_i$. However this fact alone
would not be very helpful if the correction channel itself
depended on the particular choice of channel elements.
Fortunately, it does not.

In fact, we have already exploited part of this freedom in
defining $\mathcal R$; recall that we defined it in terms of the
map
\[
\chan_\lambda(A) = \sum_i \lambda_i E_i A E_i^\dagger
\]
defined on $\mathcal B(\Hil_1)$ (see Eq.~(\ref{equ:chanlambda})).
The sequence $\lambda$ was chosen so that the infinite sum in the
expression for $\chan_\lambda$ is well defined for any operator.
However, the exact value of the components $\lambda_i$ did not
matter in the proof that $\mathcal R$ corrects the algebra
$\mathcal A$ for the channel $\chan$. The only important aspect of
this channel was that its elements are linear combinations of the
adjoints of the elements of $\chan$.

The channel $\mathcal R'$ correcting $\mathcal E'$ would be
defined in the same way in terms of the channel
\[
\begin{split}
\chan_\lambda'(A) &= \sum_i \lambda_i F_i A F_i^\dagger \\
&= \sum_{ijk} \lambda_i \gamma_{ij} \overline \gamma_{ik} E_j A E_k^\dagger\\
\end{split}
\]
which has also the right form for the corresponding correction
channel
\[
\mathcal R'(\rho) =
(\chan_\lambda')^*((\chan_\lambda'(\one))^{-\frac{1}{2}}\,
\rho\,(\chan_\lambda'(\one))^{-\frac{1}{2}})
\]
to correct the channel $\chan$. We will not go through the proof
that $\mathcal R'$ corrects $\chan$ on $\mathcal A$, since
precisely the same steps can be followed as for $\mathcal R$
itself.

Hence we have seen that  all the channels whose elements span the
same space of operators will have the same correctable algebra,
and be correctable through the same correction channel. This means
that this theory can be applied to the case described in Subsection
\ref{subsection:qec}, where the span of the elements is all that we
know about the channel.

There is a sense in which it is this fact which allows for the
quantum errors to be understood as being {\em discrete}
\cite{nielsen00}. Indeed, a standard error model for quantum
computing is that where the system considered is a tensor product
of {\em qubits}, namely two-dimensional quantum systems. The
possible ``errors'' (i.e. possible channel elements of the noise)
are supposed to be any operator acting on no more than $n$
subsystems, where $n$ is fixed. It is clear that this set of
errors is continuous. However, for a finite number of qubits their
span is separable (in fact finite-dimensional), which means that
it suffices to choose a discrete set which spans the space and try
to correct these only.

This discussion applies to simultaneously correctable sets of
observables characterized by an algebra, which are the correctable
sharp observables. However, it does not apply to the classes of
simultaneously correctable unsharp observables identified in the
previous section. Indeed, in those cases the correctable operator
systems $\mathcal S_0$ may be different for two channels whose
elements span the same operator space. To see this, remember that
\[
\mathcal S_0 = \chanh(\mathcal R_0^*(\mathcal A_0)),
\]
where $\mathcal A_0$ is the correctable algebra for the channel
restricted to a subspace $\Hil_0$, and $\mathcal R_0$ the
corresponding correction channel. Therefore, although both
$\mathcal R_0$ and $\mathcal A_0$ would be the same for both
channels, the set $\mathcal S_0$ in this expression depends
explicitly on the action of the channel itself, and may be
different in both cases.

\subsection{Nature of correctable channels}

The fact that $\mathcal A$ is the correctable algebra for the
channel $\chan$, with correction channel $\mathcal R$, implies
that the map $\chanh \circ \mathcal R^*$ acts simply as the
identity on $\mathcal A$;
\[
(\chanh \circ \mathcal R^*) |_{\mathcal A} = \id_{\mathcal A}
\]
The following theorem is a direct generalization of the results
presented in \cite{beny08x4}. It elucidates what happens to the
algebra $\mathcal A$ prior to its correction.
\begin{thm}
\label{thm:homo} Let $\mathcal A$ be the correctable algebra for a
channel $\chan$. Then $\chanh$ is a normal $*$-homomorphism of the
algebra generated by the pre-image of $\mathcal A$. In particular,
for any operators $B$, $B'$ such that $\chanh(B), \chanh(B') \in
\mathcal A$, we have
\[
\chanh(BB') = \chanh(B) \chanh(B').
\]
\end{thm}
\proof Remember that the projectors $P$ in the correctable algebra
$\mathcal A$ satisfy $B E_i = E_i P$ for some operator $B$, and
for all $i$ (see Eq.~(\ref{equ:preseff3})). It can be directly
checked that this is also true of the span of the projectors,
which is the whole of the algebra $\mathcal A$, up to closure. Now
consider two operators $B$ and $B'$ such that $A := \chanh(B)$ and
$A' := \chanh(B')$ belong to the span of the projectors of
$\mathcal A$. We know that $B E_i = E_i A$ and $B' E_i = E_i A'$.
This implies that $B B' E_i = B E_i A' = E_i A A'$, from which it
follows that
\[
\chanh(B B') = \sum_i E_i^\dagger B B' E_i = \sum_i E_i^\dagger
E_i A A' = A A' = \chanh(B)\chanh(B').
\]
Since $\chanh$ is weak-$*$ continuous, this condition also applies
to the weak-$*$ closure of the set of operators whose images are
in the span of the projectors of $\mathcal A$. This shows that the
above condition holds for every operator in the pre-image of
$\mathcal A$. \qed

Note that the pre-image of $\mathcal A$ under $\chanh$ includes in
particular the image of the dual of any correction channel
$\mathcal R$. In fact, the correction channel defined in
Eq.~(\ref{equ:R}) is itself a homomorphism. We saw in the above
proof that for all operators $A$ in the span of the projectors of
$\mathcal A$, we have $\mathcal R^*(A) E_i = E_i A$. This implies
that
\[
\mathcal R^*(A) \chan_\lambda(\one) = \chan_\lambda(A),
\]
which means explicitly
\[
(\chan_\lambda(\one))^{-\frac{1}{2}}
\chan_\lambda(A)(\chan_\lambda(\one))^{\frac{1}{2}}
=\chan_\lambda(A),
\]
or, simply,
\[
[(\chan_\lambda(\one))^{-\frac{1}{2}}, \chan_\lambda(A)] = 0.
\]
From the weak-$*$ continuity of $\chan_\lambda$, this
is true for all $A \in \mathcal A$. Using this fact, and also
recalling that $[A, E_i^\dagger E_j] = 0$ for all $A \in \mathcal
A$, yields
\[
\begin{split}
\mathcal R^*(A) \mathcal R^*(A') &= (\chan_\lambda(\one))^{-\frac{1}{2}} \chan_\lambda(A) (\chan_\lambda(\one))^{-1} \chan_\lambda(A') (\chan_\lambda(\one))^{-\frac{1}{2}} \\
&= (\chan_\lambda(\one))^{-\frac{3}{2}} \chan_\lambda(A) \chan_\lambda(A') (\chan_\lambda(\one))^{-\frac{1}{2}} \\
&= (\chan_\lambda(\one))^{-\frac{3}{2}} \chan_\lambda(\one) \chan_\lambda(AA') (\chan_\lambda(\one))^{-\frac{1}{2}} \\
&= (\chan_\lambda(\one))^{-\frac{1}{2}} \chan_\lambda(AA') (\chan_\lambda(\one))^{-\frac{1}{2}} \\
&= \mathcal R^*(A A').
\end{split}
\]
Hence, we have proved the following result.
\begin{prop}
The correction channel given by Eq.~(\ref{equ:R}) is a faithful
representation of the correctable von Neumann algebra $\mathcal
A$.
\end{prop}

This shows that the effect of a channel on its correctable algebra simply amounts to representing it in a different way on the Hilbert space. These results clarify certain aspects of \cite{kribs06x1}.  We refer to \cite{beny08x4} for the precise relationship between our results and those of \cite{kribs06x1}.

\section{Algebraic Codes}

In the rest of the paper we shall describe a number of special
cases and examples that arise in our setting.

In Subsection \ref{subsection:qec} we mentioned that the purpose of
quantum error correction was to find a ``code'' on which the
channel can be inverted, without defining what we meant by a code.
In the previous section, we started from the general assumption
that a code should be a set of simultaneously correctable
observables. We have then found that for any subspace $\Hil_0$ of
the source Hilbert space $\Hil_1$, there is a set of simultaneously
correctable observables characterized by an operator system
$\mathcal S_0$, or equivalently by the von Neumann algebra
$\mathcal A_0$ which is such that $\mathcal A_0 = V^\dagger
\mathcal S_0 V$ where $V$ is the isometry embedding $\Hil_0$ into
$\Hil_1$. The algebra $\mathcal A_0$ characterizes the sharp
observables correctable for the channel restricted to the subspace
$\Hil_0$.

Hence, all the codes that we identified, on which the channel can be inverted, are, or correspond to, von Neumann algebras. In fact it is easy to build abstract examples which yield any possible von Neumann algebra in this way, given that all von Neumann algebras can be realized as the commutant of a set of operators~\cite{davidson96}.

The general structure of von Neumann algebras is well studied. Let
us briefly review the basics here in order to understand the type
of information that they represent.

\subsection{Structure of von Neumann algebras}
\label{section:vnstruct}

Let us first summarize the representation theory of
finite-dimensional von Neumann algebras, which are just
$*$-algebras.

A concrete finite-dimensional $*$-algebra $\mathcal A$,
represented by matrices; i.e., operators on a finite-dimensional
Hilbert space represented in a fixed orthonormal basis, always has
the form
\begin{equation}
\label{equ:finalgrep} \mathcal A = \bigoplus_{k=1}^N \; \mathcal
M_{n_k} \otimes \one_{m_k}
\end{equation}
where $\mathcal M_{n_k}$ denotes the full set of square matrices of size $n_k$ and $\one_{m_k}$ the identity matrix of size $m_k$.
If the dimension of the
algebra $\mathcal A$ is $D$ then we have \( D = \sum_k n_k^2 \).
The direct sum of two matrix algebras must be understood as the
algebra of block-diagonal matrices, with one block encoding the
first algebra, and the other block the second algebra. Therefore
the above means that, written as a matrix of blocks,
\[
 \mathcal A =
 \begin{pmatrix}
    \mathcal M_{n_1}  \otimes \one_{m_1} & 0 & \cdots & 0\\
    0 & \mathcal M_{n_2}  \otimes \one_{m_2} & \cdots & 0\\
    \vdots & \vdots & \ddots & \vdots\\
    0 & 0 & \cdots & \mathcal M_{n_N} \otimes \one_{m_N}  \\
  \end{pmatrix}.
\]
In addition, tensoring a matrix algebra with the identity on
another algebra means that we are considering matrices which are
also block-diagonal, with as many blocks as there are elements on
the diagonal of the identity matrix, but such that all blocks are
identical, not only in their size, but also in their content.

For instance, any operator $A$ in the algebra $\mathcal A =
(\mathcal M_{2} \otimes \one_2) \oplus (\mathcal M_{3})$ has the
form
\[
A = \begin{pmatrix}
B & 0 & 0 \\
0 & B & 0 \\
0 & 0 & C \\
\end{pmatrix}
\]
for a two-by-two matrix $B$ and a three-by-three matrix $C$.

The block-diagonal structure of $\mathcal A$ is determined by the
form of its {\em \ind{center}} $\mathcal Z(\mathcal A)$. The
center is the set of operators inside the algebra which commute
with all other elements of the algebra;
\[
\mathcal Z(\mathcal A) = \{ A \in \mathcal A \; : \; [A,B] = 0
\text{ for all $B \in \mathcal A$ } \}.
\]
It is itself a commutative von Neumann algebra. The center can
also be written as the intersection of the algebra with its {\em
\ind{commutant}} $\mathcal A'$ which is the algebra composed of
all operators commuting with all elements of $\mathcal A$;
\[
\mathcal Z(\mathcal A) = \mathcal A \cap \mathcal A'.
\]
For instance, for an algebra of the form $\mathcal M_n \otimes
\one_m$, we have
\[
\mathcal Z(\mathcal M_n \otimes \one_m) = (\mathcal M_n \otimes
\one_m) \cap(\one_n \otimes \mathcal M_m) \approx \mathbb C.
\]
A von Neumann algebra is said to be a {\em \ind{factor}} if its
center is isomorphic to $\mathbb C$. Hence matrix algebras of the
form $\mathcal M_n \otimes \one_m$ are factors.

More generally, if the representation of $\mathcal A$ is expressed
as in Eq.~(\ref{equ:finalgrep}), then the commutant is
\[
\mathcal A' = \bigoplus_{k=1}^N \;\one_{n_k} \otimes \mathcal
M_{m_k}
\]
and the center of $\mathcal A$ is
\begin{equation}
\label{equ:fincenter} \mathcal Z(\mathcal A) = \bigoplus_{k=1}^N
\;\mathbb C (\one_{n_k} \otimes \one_{m_k})
\end{equation}
which means that it is composed of diagonal matrices with only $N$
different eigenvalues. If $P_k$ is the projector on the $k$th
block, then this means that a generic element $C \in \mathcal
Z(\mathcal A)$ of the center is of the form
\[
C = \sum_k c_k P_k
\]
for arbitrary complex numbers $c_k$. The algebra $\mathcal A$
itself is block-diagonal in terms of the subspaces defines by the
projectors $P_k$, in the sense that for all $A \in \mathcal A$,
\[
A = \sum_i P_k A P_k.
\]
Hence the center of the algebra essentially tells us what the
blocks are in its representation.

For instance, consider again the algebra $\mathcal A = (\mathcal
M_{2} \otimes \one_2) \oplus (\mathcal M_{3})$. Typical operators
$A' \in \mathcal A'$ and $C \in \mathcal A$ have the form
\[
A' = \begin{pmatrix}
a \one_2 & b \one_2 & 0 \\
c \one_2 & d \one_2 & 0 \\
0 & 0 & x \one_3 \\
\end{pmatrix}
\quad {\rm and} \quad C = \begin{pmatrix}
a \one_2 & 0 & 0 \\
0 & a \one_2 & 0 \\
0 & 0 & x \one_3 \\
\end{pmatrix},
\]
where $a,b,c,d,x \in \mathbb C$.

When $\mathcal A$ is infinite-dimensional, the direct sum must be
replaced by a {\em \ind{direct integral}}. This follows from the
fact that the center can be any commutative algebra, which has the
form
\[
\mathcal Z(\mathcal A) \approx L^\infty(\Omega)
\]
for some set $\Omega$ equipped with a measure. It is with respect
to this measure that we can write
\[
\mathcal A \approx \int^{\oplus}_\Omega \mathcal A(x) \, dx
\]
Where the generalized ``blocks'' $\mathcal A(x)$ are factors;
i.e., have a trivial center. If $\Omega$ is finite then we must
use a discrete measure, which gives us the direct sum in
Eq.~(\ref{equ:fincenter}). Hence this integral can be intuitively understood as a continuous limit of the direct sum.

Factors come in three main types. Up to now we have been using {\em \ind{type I factor}s}, which can always be represented as $\mathcal B(\Hil)$ for some Hilbert space $\Hil$. They are characterized abstractly by the fact that some of the projections they contain are {\em minimal}, which means that there is no smaller nonzero projection in the algebra. When the algebra is represented as $\mathcal B(\Hil) \otimes \one$, the minimal projections are the projectors on local pure states, i.e. of the form $P = \proj{\psi} \otimes \one$.

The factors that have no minimal projections are further
classified into types II and III according to other properties of
their projections. Type II factors have been extensively studied
in pure mathematics. Below we shall discuss a specific instance of
a type II factor that arises in our setting. Type III factors are
of central interest in physics since they describe local algebras
of observables in relativistic quantum field theory~\cite{haag93}.
In this sense they are the proper way to model local degrees of
freedom.

\subsection{Subspace codes and subsystem codes}
\label{section:oqec}

Traditionally, a {\em quantum error correcting code} is just a
subspace $\Hil_0$ of the initial Hilbert space $\Hil_1$, which is
assumed to be finite-dimensional \cite{bennett96, shor95,
steane96, gottesman96, knill97, nielsen00}. The idea is that the
channel $\chan$ is correctable for states in $\Hil_0$ if there is
a channel $\mathcal R$ such that
\[
\mathcal R(\chan(\rho)) = \rho ,
\]
for all states $\rho$ which are mixtures of pure states in the
subspace $\Hil_0$. If we introduce the isometry $V: \Hil_0
\rightarrow \Hil_1$ which embeds $\Hil_0$ into $\Hil_1$, this
means that
\begin{equation*}
\mathcal R(\chan(V \rho V^\dagger)) = V \rho V^\dagger
\end{equation*}
for all $\rho \in \mathcal B_t(\Hil_0)$, which is equivalent to
requiring the existence of a channel $\mathcal R'$ such that
\begin{equation*}
\mathcal R'(\chan(V \rho V^\dagger))  = \rho
\end{equation*}
for all $\rho \in \mathcal B_t(\Hil_0)$. Indeed, it suffices to
pick $\mathcal R'(\rho) = V^\dagger \mathcal R(\rho) V$.

If we define $\chan_0(\rho) := \chan(V \rho V^\dagger)$, this
means that $\mathcal R' \circ \chan_0$ is the identity on
$\mathcal B_t(\Hil_0)$, or equivalently that $\chanh \circ
(\mathcal R')^*$ is the identity on $\mathcal B(\Hil_0)$, which,
as we have shown matches our conception of correctability for the
algebra $\mathcal B(\Hil_0)$.  Therefore we recover the framework
of standard (subspace) quantum error correction, for a code
$\Hil_0$, when the correctable algebra is $\mathcal B(\Hil_0)$,
and the channel is restricted to the subspace $\Hil_0$.

In order to complete the comparison, let us check that our
correctability condition reduces to the one introduced for
standard codes~\cite{knill97}. The Knill-Laflamme condition states
that a standard code represented by the subspace $\Hil_0$ is
correctable for the channel $\chan$ with elements $E_i$ if and
only if there exists scalars $\lambda_{ij}\in \mathbb{C}$ such
that
\begin{equation}
\label{equ:qeccond} V^\dagger E_i^\dagger E_j V = \lambda_{ij}
\one \quad \forall i,j,
\end{equation}
where $V$ embeds $\Hil_0$ into the source Hilbert space $\Hil_1$.
In our framework, the correctable algebra must be precisely the
commutant of the operators $V^\dagger E_i^\dagger E_j V$ for all
$i$ and $j$. Since here they are all proportional to the identity
on $\Hil_0$, the correctable algebra is indeed the whole algebra
of operators on $\Hil_0$.

A more general framework was also introduced which generalized the
notion of a code to that of a {\em subsystem code} \cite{kribs05,
kribs06}. In this approach one defines a code through a subspace
$\Hil_0 \subseteq \Hil_1$ (where $\Hil_1$ is still
finite-dimensional) and a particular subsystem decomposition
$\Hil_A \otimes \Hil_B = \Hil_0$ of this subspace. Again, let $V$
be the isometry embedding $\Hil_0$ into $\Hil_1$. We then say that
the subsystem $\Hil_A$ is a correctable code if there is a channel
$\mathcal R$ such that
\[
\mathcal R(\chan(V(\rho \otimes \tau) V^\dagger)) = \rho \otimes
\tau'
\]
for any states $\rho \in \mathcal B_t(\Hil_A)$, $\tau, \tau' \in
\mathcal B_t(\Hil_B)$.  We want to show that this is equivalent to
the case where the correctable algebra $\mathcal A$, in our
framework, is any finite-dimensional factor of type I, which in
this case is
\[
\mathcal A = \mathcal B(\Hil_A) \otimes \one_B.
\]
That is, assuming we are restricting the initial state to the
subspace $\Hil_0$. In our language, this would mean that
\begin{equation}
\label{equ:vexev} V^\dagger \chanh(\mathcal R^*( X\otimes \one ))
V = X \otimes \one
\end{equation}
for all $X \in \mathcal B(\Hil_A)$. Indeed, suppose first that
$\Hil_A$ is a subsystem code corrected by $\mathcal R$ , then we
have for all $X \in \mathcal B(\Hil_A)$,
\[
\begin{split}
\tr(V^\dagger \chanh(\mathcal R^*(X\otimes \one)) V
(\rho\otimes\sigma))
&= \tr((X\otimes \one) \mathcal R (\chan(V \rho \otimes \sigma V^\dagger)))\\
& = \tr(X\rho \otimes \tau) = \tr(X \rho) \tr(\tau) \\
&= \tr(X \rho) = \tr((X \otimes \one) (\rho \otimes \sigma)).
\end{split}
\]
This is true for all states $\rho \in \mathcal \mathcal
B_t(\Hil_A)$ and all states $\sigma \in \mathcal B_t(\Hil_B)$. By
linearity it follows that $V^\dagger \chanh(X\otimes \one)V = X
\otimes \one$ for all $X \in \mathcal \mathcal B(\Hil_A)$.
Conversely, if Eq.~(\ref{equ:vexev}) is true for all $X$, then for
all $\rho \in \mathcal \mathcal B_t(\Hil_0)$ we have
\[
\begin{split}
\tr(X\, \tr_B(\mathcal R(\chan( V \rho V^\dagger))))
&= \tr( (X\otimes \one) \mathcal R(\chan(V \rho V^\dagger))) \\
&= \tr(V^\dagger \chanh(\mathcal R^*(X\otimes \one)) V \rho ) \\
&= \tr((X\otimes \one) \rho ) \\
&= \tr(X \,\tr_B(\rho)).
\end{split}
\]
Since the above equation is true for all $X$, we have
$\tr_B(\mathcal R(\chan( V \rho V^\dagger))) = \tr_B(\rho)$ for
all $\rho \in \mathcal B_t(\Hil_0)$, which was shown in
\cite{kribs06} to be equivalent to the definition of $\Hil_A$
being a noiseless subsystem for $\chan$.

In this framework, the correctability condition reads
\cite{kribs05}
\begin{equation}
\label{equ:oqeccond} V^\dagger E_i^\dagger E_j V = \one \otimes
\Lambda_{ij}
\end{equation}
for an arbitrary set of operators $\Lambda_{ij} \in \mathcal
B(\Hil_B)$.  This means that the operators $V^\dagger E_i^\dagger
E_j V$ for all $i$ and $j$ generate the sub-algebra $\one \otimes
\mathcal B(\Hil_B)$ of $\mathcal B(\Hil_0)$, whose commutant is
indeed $\mathcal B(\Hil_A) \otimes \one$; the correctable algebra
defining the subsystem code.

Note that every subsystem code is associated with a family of
standard codes, whenever we can afford to put stronger constraints
on the initial state. Indeed, consider the smaller subspace
$\Hil_0'$ formed by the states inside $\Hil_0$ which are of the
form $\ket{\psi} \otimes \ket{\phi_0}$, where $\ket{\phi_0}$ is
fixed. This subspace is associated with the isometry $W = V
\otimes \ket{\phi_0}$, for which we have
\[
W^\dagger E_i^\dagger E_j W = \bra{\phi_0} \Lambda_{ij}
\ket{\phi_0} \one
\]
which is just the Knill-Laflamme condition for $\Hil_0'$.

These results show that the standard subspace codes, as well as
the subsystem codes, correspond to the case where our correctable
algebra is a finite-dimensional factor, which is always of type I.
Our results yield several types of generalizations over these codes.
Firstly, we obtain a characterization of infinite-dimensional
quantum codes, which will be
discussed further in Subsection~\ref{section:infdimcodes} and
Sections~\ref{section:extypeone} and \ref{section:extypetwo}.
We also obtain a characterization of discrete or continuous classical codes, corresponding to a commutative correctable algebra. This case will not be discussed here.
In addition, we can handle the  correction of information that is
neither quantum nor classical; i.e., that is represented by an
algebra which is neither commutative nor a factor.

\subsection{Hybrid classical-quantum codes}
\index{hybrid codes}

We have seen in our framework that the structure to be corrected
can be any von Neumann algebra. A general von Neumann algebra with
center $\mathcal Z(\mathcal A) = L^\infty(\Omega)$ is of the form
\begin{equation}
\label{equ:hybrid} \mathcal A = \int_\Omega^\oplus \mathcal A(x)
dx
\end{equation}
where each $\mathcal A(x)$ is a factor. If the center is maximal,
$\mathcal Z(\mathcal A) = \mathcal A$, then the algebra $\mathcal
A$ is commutative and each factor $\mathcal A(x)$ is of dimension
one, and hence isomorphic to the complex numbers $\mathbb C$. If,
on the other hand, the center is minimal, $\mathcal Z(\mathcal A)
\approx \mathbb C$, then the set $\Omega$ contains only a single
element $x_0$ and $\mathcal A = \mathcal A(x_0)$ is a factor.

If $\mathcal A$ is commutative, then it represents a classical
system, which is clear from the fact that it has the form
$L^\infty(\Omega)$. It is then natural to say that if it is a
factor, it represents a ``pure'' quantum system.

A physical system represented by an algebra, whose structure is
given by the general form represented in Eq.~(\ref{equ:hybrid}), can be
understood as being partly quantum and partly classical. Indeed,
we can consider the center $\mathcal Z(\mathcal A) =
L^\infty(\Omega)$ as representing a classical system. For each
possible ``state'' $x \in \Omega$ of this classical system, we
have a pure quantum system represented by the factor $\mathcal
A(x)$.

For instance, a classical system, represented by
$L^\infty(\Omega)$, next to a type I quantum system, with algebra
$\mathcal B(\Hil)$, is represented by
\[
\mathcal A = L^\infty(\Omega) \otimes \mathcal B(\Hil) \simeq
\int^\oplus_\Omega \mathcal A(x) dx
\]
where each factor $\mathcal A(x)$ is a copy of $\mathcal B(\Hil)$.
In the more general case, however, the size and type of the
algebra $\mathcal A(x)$ may depend upon $x$.

Generic operators $A,B \in \int_\Omega^\oplus \mathcal A(x) \, dx
$ are of the form
\begin{equation}
\label{equ:hybrideffect} A = \int_\Omega A_x \, dx \quad\quad
\text{and}\quad\quad B = \int_\Omega B_x \, dx ,
\end{equation}
where $A_x, B_x \in \mathcal A(x)$ for all $x \in \Omega$. Their
product is simply
\[
A B = \int_\Omega A_x B_x \, dx.
\]
An element of the center is of the form
\[
C = \int_\Omega \alpha(x) \one_x\, dx
\]
where $\alpha \in L^\infty(\Omega)$ and $\one_x$ is the identity
on $\mathcal A_x$.

If $A$ in Eq.~(\ref{equ:hybrideffect}) is an effect, then each
operator $A_x$ is also an effect that can be interpreted as a
quantum proposition which is true conditionally on the classical
system being in state $x$.

If the finite-dimensional case, if a hybrid algebra
\[
\mathcal A = \bigoplus_k \mathcal M_{n_k} \otimes \one_{m_k}
\]
is correctable, then each factor $\mathcal A_k = \mathcal M_{n_k}
\otimes \one_{m_k}$ represents a correctable subsystem code for
states restricted to the subspace $\Hil_k$ projected onto by $P_k
= \one_{n_k} \otimes \one_{m_k}$. Indeed, if $V$ is the isometry
corresponding to $P_k$, then $V A V^\dagger$ is the representation
inside $\mathcal A$ of an operator $A \in \mathcal A_k$. Thus
\[
V^\dagger \chanh(\mathcal R^*(V A V^\dagger)) V = V^\dagger V A
V^\dagger V = A.
\]
Hence a finite-dimensional hybrid algebra can be understood
as representing a family of orthogonal subsystem codes correctable
simultaneously.

\subsection{Infinite-dimensional subspace and subsystem codes}
\label{section:infdimcodes}

The derivation of necessary and sufficient conditions for error
correction of infinite-dimensional algebras is a new result which
is potentially significant, given that all physical systems are
naturally modelled by infinite-dimensional systems. In particular,
it yields a formulation of quantum error correction for systems
characterized by continuous variables \cite{braunstein03}.

As we discussed above, a code can be said to be purely quantum if
it is represented by an algebra which is a factor. In the
finite-dimensional case, we have seen in
Subsection~\ref{section:oqec} that factors represent subsystem
codes characterized by Eq.~(\ref{equ:oqeccond}), or
Eq.~(\ref{equ:qeccond}). Some authors assumed that this condition
would hold unchanged in the infinite-dimensional case. For
instance, in \cite{braunstein98x1} the Knill-Laflamme condition
was expressed for a channel with continuous elements $E_x$ as
\[
V^\dagger E_x^\dagger E_y V = \lambda(x,y) \one.
\]
where $\lambda(x,y) \in \mathbb C$.
Our results show that this condition is in fact \emph{sufficient}.
Indeed, it implies that the commutant of the operators $V^\dagger
E_x^\dagger E_y V$ for all $x$ and $y$ is the whole algebra
$\mathcal{B}(\Hil_0)$ on the subspace $\Hil_0 \subseteq \Hil_1$.
However, in the infinite-dimensional case this condition is no
longer necessary since it expresses that the code must be
isomorphic to $\mathcal{B}(\Hil_0)$, which is a factor of type I.
Hence it misses the possibility of correcting more general
factors. In particular, our generalization of quantum error
correction to infinite-dimensional systems introduces new types of
quantum codes not previously considered, namely
infinite-dimensional type I subsystem codes, factors of type II,
and factors of type III. An example of a correctable type II
factor is given in Section \ref{section:extypetwo} below.

In the next section we illustrate our framework with an elementary example of an infinite-dimensional type I factor code.

\section{Type I Infinite-Dimensional Example}
\label{section:extypeone}

Here we consider what is arguably the simplest class of bona fide
infinite-dimensional quantum error correcting subspace codes.

Let $\mathcal{H}_0$ be a subspace of an infinite-dimensional
Hilbert space $\mathcal{H}$. The only constraint we place on the
dimensionality of $\mathcal{H}_0$ is that it have infinite
co-dimension (in particular it could be infinite-dimensional
itself).


Let $P$ be the projection of $\mathcal{H}$ onto $\mathcal{H}_0$,
and let $\{P_i\}_{i=1}^\infty$ be a family of projections of
$\mathcal{H}$ onto mutually orthogonal subspaces $\mathcal{H}_i$,
each of the same dimension as $\mathcal{H}_0$. Further let
$\{p_i\}_{i=1}^\infty$ satisfy $p_i > 0$ and $\sum_{i\geq 1} p_i
=1$, and let $\{V_i\}_{i=1}^\infty$ be a family of isometries with
$\mathcal{H}_0$ as their common initial space such that $V_i$ maps
$\mathcal{H}_0$ isometrically onto $\mathcal{H}_i$. This is
equivalent to the following operator equations:
\[
P_iP_j=\delta_{ij}P_i, \quad V_i^\dagger V_i = \one_{\Hil_0}, \quad V_i
V_i^\dagger = P_i, \quad V_i = P_i V_i, \quad \forall i,j.
\]

It follows that $\mathcal{H}_0$ is entirely correctable for any channel $\chan$ which, when restricted to $\Hil_0$, acts as
\[
\chan_0(\rho) = \sum_{i= 1}^\infty p_i \, V_i \rho
V_i^\dagger.
\]
Indeed, since $V_i^\dagger V_j = \delta_{ij} \one_{\Hil_0}$, the commutant of the set $\{V_i^\dagger V_j\}_{ij}$ is the whole of $\mathcal{B}(\mathcal{H}_0)$. This class of channels and codes
is a straightforward generalization of the class that plays a
central role in the proof of the Knill-Laflamme characterization
\cite{knill97} of subspace codes for the finite-dimensional case.

For the sake of the example let us compute the correction channel
by directly applying Eq.~(\ref{equ:R}). First note that $\chan_0$
is well defined on any bounded operator, therefore we can use it
in place of the modified channel $\chan_\lambda$ introduced in Eq.~\ref{equ:chanlambda}.
We have
\[
(\chan_0(\one))^{-\frac{1}{2}} = \sum_i \frac{1}{\sqrt{p_i}} P_i ,
\]
so that the correction channel is
\[
\begin{split}
\mathcal R(\rho) = \sum_{ijk} p_i V_i^\dagger \frac{1}{\sqrt{p_j}} P_j \rho P_k \frac{1}{\sqrt{p_k}} V_i
= \sum_i V_i^\dagger \rho V_i.
\end{split}
\]
This illustrates how the correction channel becomes independent of
the particular probabilities $p_i$.

\section{Type II Example: Irrational Rotation Algebra}
\label{section:extypetwo}

Consider the algebra generated by two elements $\hat x$ and $\hat
p$ satisfying the canonical commutation relations
\[
[\hat x, \hat p] = i \one.
\]
This algebra can be represented on $\Hil = L^2(\mathbb R)$, where
the position operator $\hat x$ acts on a function $\psi \in
L^2(\mathbb R)$ as $(\hat x \psi)(x) = x \psi(x)$ and the momentum
$\hat p$ as $(\hat p \psi)(x) = i \frac{d}{dx} \psi(x)$.

Suppose that this system interacts with an environment through a
Hamiltonian of the form $H = \sum_i J_i \otimes K_i$, where the
operators $J_i$ act on the system, and the operators $K_i$ on the
environment.  We further assume that the interaction operators $J_i$ are of
two forms: some are periodic functions of $\hat x$, with period
$L_x$, and others are periodic functions of $\hat p$, of period
$L_p$.
This implies that these functions are linear combinations
of powers of the functions $x \mapsto e^{i \frac{2 \pi}{L_x} \,x}$
or $p \mapsto e^{i \frac{2 \pi}{L_p} \,p}$ respectively (their
discrete Fourier components). For convenience, let us define
\[
\omega_x := \frac{2 \pi}{L_x} \quad \text{ and } \quad \omega_p :=
\frac{2 \pi}{L_p}.
\]
Since the interaction operators are bounded, we can follow the reasoning presented in Subsection \ref{subsection:qec} and conclude that the channel elements of the resulting channel on the system belong to the algebra generated by the operators $J_i$.

The von Neumann algebra generated by the interaction operators is
also generated by the two unitary operators
\[
U = e^{i \omega_x \,\hat x} \quad \text{and} \quad V = e^{i
\omega_p \, \hat p} .
\]
These can be understood as the two possible errors in our noise model.

To make things more interesting, we assume that the real number
\[
\theta := \frac{\omega_x \omega_p}{2\pi}
\]
is irrational. This number is important because it enters into the
commutator of $U$ and $V$:
\[
UV = e^{2 \pi i \theta} V U.
\]

For this example, we will look for codes which are correctable for any initial state. In principle, in order to find the correctable algebra, we need to find the operators commuting with the products $E_i^\dagger E_j$
of the channel elements $E_i$. However, since we only know the
span of these operators, we cannot exclude that $E_i = \one$ for
some $i$. If this is the case, then these products include $E_i$
and $E_i^\dagger$ for all $i$. Therefore, for an effect to be
correctable, it needs to be in the commutant of the von Neumann
algebra generated by the operators $E_i$ for all $i$, which is the
same as the von Neumann algebra generated by the interaction
operators $J_i$, or simply $U$ and $V$.

The operators
\[
U' =  e^{i \frac{\omega_x}{\theta} \,\hat x} \quad \text{and}
\quad V' =  e^{i \frac{\omega_p}{\theta} \,\hat p}
\]
commute with both $U$ and $V$. 
To see that $U'$ commutes with $V$, simply note that
\[
U' V = e^{i \frac{\omega_x \omega_p}{\theta} } V U' =  e^{i\frac{2\pi \theta}{\theta}} V U' = V U'.
\]
Similarly, $V'$ also commutes with both $U$ and $V$. In fact, the
von Neumann algebra generated by $U'$ and $V'$ is the whole
commutant of the algebra generated by the interaction operators
$J_i$. In addition, it happens to be a factor of
type II, and, together, with its commutant they generate the whole
of $\mathcal B(\Hil)$ \cite{faddeev95}.

Therefore, this is an example of a correctable factor of type II.
In fact, it is also {\em noiseless} \cite{knill00}, in the sense
that the correction channel can be taken to be the identity
channel; i.e., no active correction is needed. This happens simply
because $\one$ was assumed to be among the channel elements.
Indeed, we saw that it implied that the correction operators had
to commute with the channel elements themselves. Hence, if $\chan$
is the channel, we have $\chanh(A) = \chanh(\one) A = A$ for all
elements $A$ of the correctable algebra.

Let us see how we can understand this ``\ind{type II subsystem}'',
and how it resembles, and differs from, the factors of type I with
which we are familiar.

If we were dealing with a factor of type I
containing the identity, then the Hilbert space would take the
form $\Hil = \Hil_1 \otimes \Hil_2$, so that our algebra would be
simply $\mathcal B(\Hil_1) \otimes \one$.  For
instance, consider $\Hil = L^2(\mathbb R^2) = L^2(\mathbb R)
\otimes L^2(\mathbb R)$. If $\psi \in L^2(\mathbb R^2)$, the
operators in the first factor $\mathcal A = \mathcal B(L^2(\mathbb
R)) \otimes \one$ are those which act only on the first component
of $\psi$. The first factor is generated by the
operators $(\hat x \psi)(x,y) = x\psi(x,y)$ and $(\hat p
\psi)(x,y) = \frac{d}{dx}\psi(x,y)$. Note that here the set
$\mathbb R^2$ on which the states are defined can be understood to
be the set of joint eigenvalues to the position operators in
$\mathcal A$ and $\mathcal A'$.

Something similar happens for our factor of type II. Let $\mathcal
A$ be the factor generated by $U$ and $V$, and $\mathcal A'$ its
commutant, which is generated by $U'$ and $V'$. Let us see if we
could see the elements of $\Hil$ as wavefunctions over the
eigenvalues of $U$ and $U'$. First, note that the spectrum of both
these operators is the unit circle in the complex plane. These two
operators being functions of the position operator $\hat x$, we
may wish to use the fact that the states of $\Hil$ can be
represented as wavefunctions over the spectrum of $\hat x$, that
is elements of $L^2(\mathbb R)$. Indeed, we can naturally convert
an eigenvalue $x$ of $\hat x$ into the eigenvalues
\begin{equation}
\label{eq:extworep} a = e^{i\omega_x x} \quad \text{ and } \quad b
= e^{i \frac{\omega_x}{\theta} x }
\end{equation}
respectively of $U$ and $U'$. In fact, this relationship is invertible. Indeed, if we are given $a$ and $b$, then only a single real number $x$ will satisfy both these relations. Indeed, suppose that we had two different real numbers $x$ and $x'$ yielding the same values of $a$ and $b$. This would imply that they are related by $x-x' = 2 \pi n /\omega = 2 \pi m \theta /\omega$ for two integers $n$ and $m$. But this would imply $\theta = n/m$, which is not possible since we assumed $\theta$ to be irrational.
This implies that for each state $\psi
\in L^2(\mathbb R)$, we can define the function
\[
\widetilde \psi(a,b) := \psi(x)
\]
where $x$ is the unique real number related to $a$ and $b$ via Eq.~(\ref{eq:extworep}).
Note that this function $\widetilde \psi$ is defined only on the valid couples $(a,b)$ related to some $x \in \mathbb R$ via Eq.~\ref{eq:extworep}. However, due to the irrationality of $\theta$, these couples are dense in the unit torus. We can therefore think of $\widetilde \psi$ as being defined almost everywhere.

We will see that $\widetilde \psi$ can be interpreted, in a suitable sense, as the wavefunction of a particle on a two-dimensional \ind{torus}.

The relation between the wavefunctions $\widetilde \psi(a,b)$ and
the factors $\mathcal A$ and $\mathcal A'$ is given by the fact
that they ``act'' respectively on the first and second arguments
of $\psi$ respectively. Specifically, observe that
\[
(U \widetilde \psi)(a,b) = (U \psi)(x) = e^{i\omega_x x} \psi(x) =
a \, \widetilde \psi(a,b),
\]
which means that $U$ acts just like the first component of the
position of the particle. Similarly,
\[
(U' \widetilde \psi)(a,b) = (U' \psi)(x) =
e^{i\frac{\omega_x}{\theta} x} \psi(x) = b \, \widetilde
\psi(a,b).
\]
The action of $V$ is also easy to compute,
\[
(V \widetilde \psi)(a,b) = (V \psi)(x) = \psi(x + \omega_p) =
\psi(a \,e^{2 \pi i \theta}, b).
\]
Hence, its action is to rotate the first argument by the
irrational angle $\theta$. Similarly,
\[
(V' \widetilde \psi)(a,b) = (V' \psi)(x) = \psi(x +
\omega_p/\theta) = \psi(a , b\,e^{2 \pi i \frac{1}{\theta}}).
\]

Although it looks like a particle on the torus, this system
differs from it by the nature of the normalized states. Indeed,
the norm is
\[
\|\widetilde \psi\|^2 = \int |\psi(x)|^2 \, dx = \int | \widetilde
\psi( e^{i\omega_x x}, e^{i \frac{\omega_x}{\theta} x }) |^2 \,dx
,
\]
and so what we have done is to take a standard particle in a
one-dimensional space, and wrap its space around a torus in a
dense trajectory. If we view the particle as a wavefunction
$\widetilde \psi$ on the torus, its norm is an integral over this
path which is dense in the torus.

This picture illustrates what the noise does. It disturbs only
the first component of the position of this particle, but not the
second.

\section{Outlook}
\label{section:outlook}

There are several significant lines of investigation that are
suggested by this work. One is to use the theory presented here as
the basis to extend more of the results of quantum error correction
to the infinite-dimensional setting. While some results and
techniques will surely extend straightforwardly, the present results
indicate a range of qualitatively new phenomena and correspondingly
new possibilities for quantum error correction in the
infinite-dimensional case, such as correctable type II and type III
factors. Type III factors, in particular, naturally describe local
algebras of observables in relativistic quantum field
theory~\cite{haag93}. We close by briefly touching on three specific
potential avenues of research.

First, in this work we have assumed that the noise model was given
by a quantum channel specified by the span of its elements, $E_k$.
Alternatively, a noise model can be specified by giving the
interaction Hamiltonian. The link between the two descriptions
(described for the case of bounded interaction operators in
Subsection~\ref{subsection:qec}), becomes nontrivial in the
infinite-dimensional setting. For instance, whereas the unitary
time evolution operator is bounded, the interaction operators are
generally unbounded and have a restricted domain and range.

Second, we note that although we have characterized a large class of
simultaneously correctable unsharp observables (POVM)
(Section~\ref{section:simultcorr}), it is not clear as yet whether
or not this exhausts all possible correctable observables. Previous
results~\cite{blume-kohout08} indicate that there are no other
correctable observables in finite-dimensional Hilbert spaces, but
these results need to be generalized to the infinite-dimensional
setting.

Finally, it should be interesting and relatively straightforward to
extend our results to more general channels. For instance, it would be natural to consider normal
completely positive maps defined between arbitrary von Neumann
algebras, rather than just type I factors.

In conclusion, there are clearly deep connections between quantum
error correction on the one hand and the rich field of operator
algebras and operator systems on the other, leading to new
possibilities for fruitful interactions between  the two
disciplines.


\vspace{0.6in}

\noindent{\it Acknowledgements.} Part of this work is contained in
the first named author's doctoral thesis~\cite{beny08x5}. The
authors were partly supported by the Discovery Grant, Discovery
Accelerator Supplement, and Canada Research Chair programs of NSERC,
by Ontario Early Researcher and PREA Awards, and by CFI and OIT. The
Centre for Quantum Technologies is funded by the Singapore Ministry
of Education and the National Research Foundation as part of the
Research Centres of Excellence programme.

\vspace{0.8in}




\bibliographystyle{habbrv} 

\bibliography{pig}


\end{document}